\documentclass{PoS}
\usepackage{amssymb,amsmath,graphicx}

\title{Nucleon and Pion Form Factors from $N_f=2+1$ Anisotropic Lattices\footnote{NT@UW-11-07}}

\ShortTitle{$N_f=2+1$ Nucleon and Pion Form Factors}

\author{Huey-Wen Lin\\
Department of Physics, University of Washington, Seattle, WA 98195-1560 \\
}

\author{Saul D. Cohen \\Center for Computational Science, Boston University, Boston, MA 02215
}

\abstract{
We report a recent lattice-QCD calculation of nucleon and pion electromagnetic form factors and nucleon axial form factors, with special emphasis on large $Q^2$. Conventional lattice form-factor calculations can only reach about 2.5~GeV$^2$, but in this work the transfer momentum is pushed as large as 6~GeV$^2$. Here, we demonstrate the results on 2+1-flavor anisotropic clover lattices for the nucleon and pion, comparing with low-$Q^2$ quantities, such as Dirac and Pauli radii, anomalous magnetic moments, $g_A$ and $M_A$. Our approach can be applied to isotropic lattices and lattices with smaller lattice spacing to achieve even larger-$Q^2$ form factors. The form factors are processed to obtain transverse charge and magnetization densities across 2-dimensional impact-parameter space.
These measurements could give important theoretical input to experiments, such as those of JLab's 12-GeV program, and provide insight into hadronic structure.
}

\FullConference{The XXVII International Symposium on Lattice Field Theory\\
June 14-19, 2010 \\
Villasimius, Sardinia\\
}

\begin{document}

\section{Introduction}
Nucleons are the building blocks of ordinary matter and pions are the lightest hadrons existing in our world. In the Standard Model, they are composed from fundamental particles, quarks and gluons; non-trivial interactions of quarks and gluons within the confined space of hadrons determine the properties of these hadrons. Due to  confinement, we cannot directly observe how the quarks and gluons compose the pion and nucleon; rather, experiments smash hadrons into different final states (as in deep inelastic scattering) to provide us with understanding through observables, such as the parton distribution functions (PDFs). Processes such as electron elastic scattering with hadrons via photons give us electromagnetic form factors, which contain information in the momentum-space impact plane. Adding up information from various experiments, we can reconstruct these processes to gain a better understanding of how the quarks and gluons shape the nucleon and pion.

Understanding form factors at high transfer momentum ($Q^2$) is particularly interesting. It not only helps to reveal the nature of hadrons at short distances but also challenges models whose parameters are calibrated in the low-$Q^2$ regime.
Experiments such as those in the 12-GeV upgrade at Jefferson Lab will soon provide higher precision and higher momentum-transfer form-factor data; progress will be made not only in the nucleon sector but also for pions.
Theoretically, perturbative QCD (pQCD) should perform better at larger $Q^2$, and we would like to get an understanding of when pQCD starts to become reliable. However, pQCD seems to fail to describe recent BaBar results for $\gamma^*\gamma \rightarrow \pi^0$ over a wide range of transfer momenta $4\mbox{ GeV}^2 < Q^2 < 40\mbox{ GeV}^2$~\cite{:2009mc,Radyushkin:2009zg}, suggesting that there may be a large gap before pQCD can describe the experimental data correctly. A nonperturbative QCD approach would be able to provide the QCD theory results to connect the missing dots if one could use such an approach for sufficiently large $Q^2$.

The techniques of lattice QCD have been applied to such varied phenomena as the spectroscopy of heavy-quark hadrons, the tower of excited baryon states, flavor physics involving the CKM matrix, hadron decay constants and baryon axial couplings. In many cases, it can provide higher-precision data from the Standard Model than what can be measured experimentally. 
By discretizing spacetime into a four-dimensional grid with a fixed lattice spacing and finite volume, we are able to compute the path integral (in terms of discretized versions of the QCD Lagrangian and operators) directly via numerical integration, providing first-principles calculations of the consequences of QCD. The four-dimensional lattice size, lattice spacing, the quark masses in the sea sector and precision of the integration have been greatly improved due to the increasing computational resources and more efficient algorithms over the past decade. Preliminary calculations at physical pion masses are being explored. Lattice calculations with sub-percent precision have been achieved for many flavor-related quantities and used to guide tests of Standard Model and to probe possible new physics. Calculations of hadron structures and interactions require more computational resources but should soon enter an era of precision.

Hadron form factors have been calculated on the lattice by many groups, and calculations are still ongoing; see Refs.~\cite{Alexandrou:2010cm,Hagler:2009ni,Zanotti:2008zm,Orginos:2006zz} and references within for a few examples of recent review articles. Recently, lattice calculations have also been used to calculate nucleon transition form factors involving excited nucleons~\cite{Lin:2008qv} and experimentally challenging quantities, such as the hyperon axial couplings~\cite{Lin:2007ap}, for the first time. However, the typical $Q^2$ range in lattice calculations of hadron form factors is less than $2.5\mbox{ GeV}^2$ on a typical lattice spacing $a\approx 0.12$~fm. When one attempts higher-$Q^2$ calculations, they suffer from poor signal-to-noise ratio; for example, see a case study for the pion in Ref.~\cite{Hsu:2007ai}.

There are two main obstacles in lattice QCD to achieving higher-$Q^2$ form factors. One limitation is the lattice spacing ($a$) available to be used in the calculation, since the momenta allowed on the lattice (assuming periodic spatial boundary conditions) are $\frac{2\pi n}{L}a^{-1}$, where $L$ is the length of the spatial volume while $n$ ranges 0, 1, $\sqrt{2}$, $\sqrt{3}$, etc. Thus, the finer the lattice spacing, the higher momentum transfer one can achieve. However, very fine lattices may not be available in the next couple years and the spatial volume needs to be sufficiently large to avoid finite-volume effects, so we must instead increase the momentum in lattice units.
In many existing calculations, the signal-to-noise ratio in form factors becomes quite poor before reaching high $n$ at a given lattice spacing. We have examined the typical steps in a form-factor calculation to find the main causes of this noise and to develop solutions. The traditional method simplifies the complicated form-factor analysis by working in a regime where only the ground-state signal remains. This greatly simplifies the analysis, since taking ratios of the three-point correlator with current-insertion and the hadron two-point correlators will yield a constant in Euclidean time, which can be fit to a simple plateau. To suppress excited states in the limited time extent of the lattice box, we usually use interpolating operators that have broad extent in space, called ``smeared''; this gives the same ground-state signal at earlier Euclidean time. The parameters of the smearing are usually tuned by looking at the hadron masses at rest or checking the form factors at zero or very low momentum transfer. However, as we increase the magnitude of the momentum transfer carried by the currents, the momentum carried by the initial or final hadron state must also increase (because of momentum conservation). The smearing parameters used to filter out the excited states will also filter out the boosted hadron signal. Thus, in a system carefully tuned to include only the ground state, the signal will die out after a few momentum points 
when using fixed smearing parameters throughout the calculation. The solution to this problem is simple: we can vary the smearing parameters as the current momentum changes but keep a close eye on the contamination from excited states for each momentum; or we can take not only the ground state but also the excited state into the analysis, extracting the ground state using multiple smearing parameters or operators that has good overlap with various different states.

We first investigated this idea using a quenched lattice with pion mass 720~MeV, calculating the $d$-quark connected contribution to nucleon form factors. For this demonstration, we fixed the final-state nucleon at rest so that the initial nucleon would carry the momentum transfer out at the inserted current. Ref.~\cite{Lin:2010fv} shows the results from both methods: In the low-momentum region, we find the two approaches show consistent results with compatible statistical error bars.
However, as soon as the momentum increases above the rest excited-state energy set by the smearing parameter, the initial nucleon operator contributes no further signal and presents a great deal of noise. At that point, some of our parameters that couple well to the rest excited state become dominant for the boosted nucleon ground state. These contribute to the form factors in the new method, and the signal-to-noise ratio remains relatively stable.
Therefore, we can reach higher momentum transfer than the traditional fixed-operator, ground-state-only approach.

In this proceeding, we will demonstrate the results of this improved method, focusing on results from dynamical lattices. 
We will not only cover nucleon electromagnetic form factors, which are more commonly calculated and better covered by experimental data over $Q^2$, but also the less known large-$Q^2$ region for nucleon axial and pion form factors. At the end of this proceeding, we will propose a step-scaling method for a future calculation of form factors at even higher momentum transfer, free of the systematics introduced in a calculation at fixed lattice spacing.

\section{Numerical Results}

The main body of our work uses the 2+1-flavor anisotropic lattices generated by the Hadron Spectrum Collaboration (HSC)~\cite{Edwards:2008ja,Lin:2008pr}. (Note that the same techniques will work equally well for isotropic lattices.)
These lattices use Symanzik-improved gauge action with tree-level tadpole-improved coefficients and anisotropic clover action~\cite{Chen:2000ej}. The gauge links in the fermion action are 3-dimensionally stout-link smeared with a small smearing weight $\rho=0.14$ and $n_\rho=2$ iterations. The renormalized gauge and fermion anisotropies are around $\xi=3.5$ (that is, $a_s=3.5 a_t$), and the inverse of the spatial lattice spacing is about 1.6~GeV. 
The quark propagators are calculated under antiperiodic boundary conditions in the time direction, while the spatial ones remain periodic.
Quark propagators on the lattices are evaluated for source and sink operators with five Gaussian smearing parameters: $\sigma \in \{0.5, 1.5, 2.5, 3.5, 4.5\}$. 
We use four final momenta ($\vec{p_f}=\frac{2\pi}{L}\{0,0,0\}a^{-1}$, $\frac{2\pi}{L}\{-1,0,0\}a^{-1}$, $\frac{2\pi}{L}\{-1,-1,0\}a^{-1}$, $\frac{2\pi}{L}\{-2,0,0\}a^{-1}$) and vary the initial momentum over all $\vec{p_i}=\frac{2\pi}{L}\{n_x,n_y,n_z\}a^{-1}$ with integer $n_{x,y,z}$ and $n_x^2+n_y^2+n_z^2 \leq 10$ for current matrix elements. 
For more details on the parameters used in our work, please see the Ref.~\cite{Lin:2010fv}.

\subsection{Nucleon Dirac and Pauli Form Factors}

\begin{figure}
\begin{center}
\includegraphics[width=0.48\textwidth]{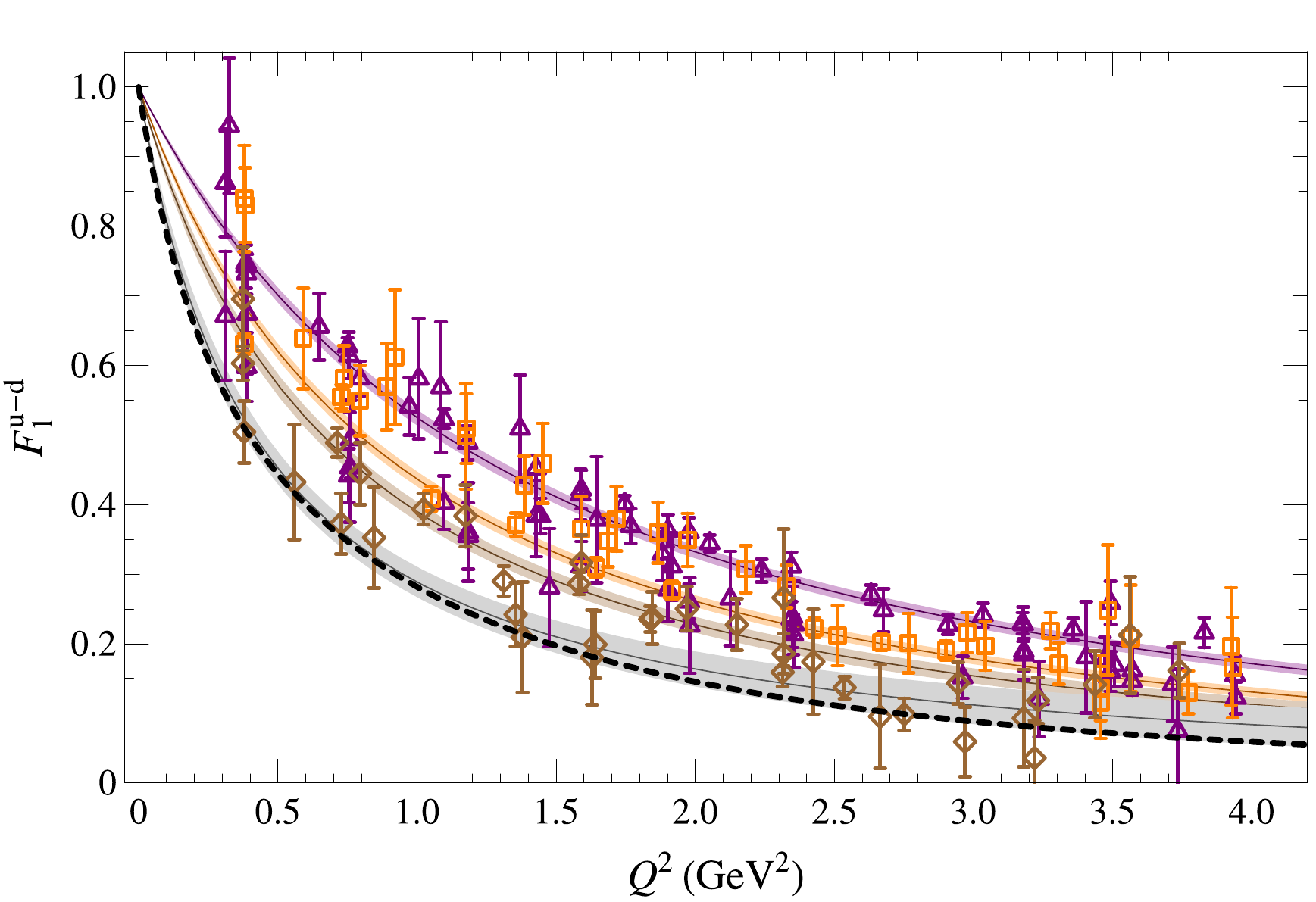}
\includegraphics[width=0.48\textwidth]{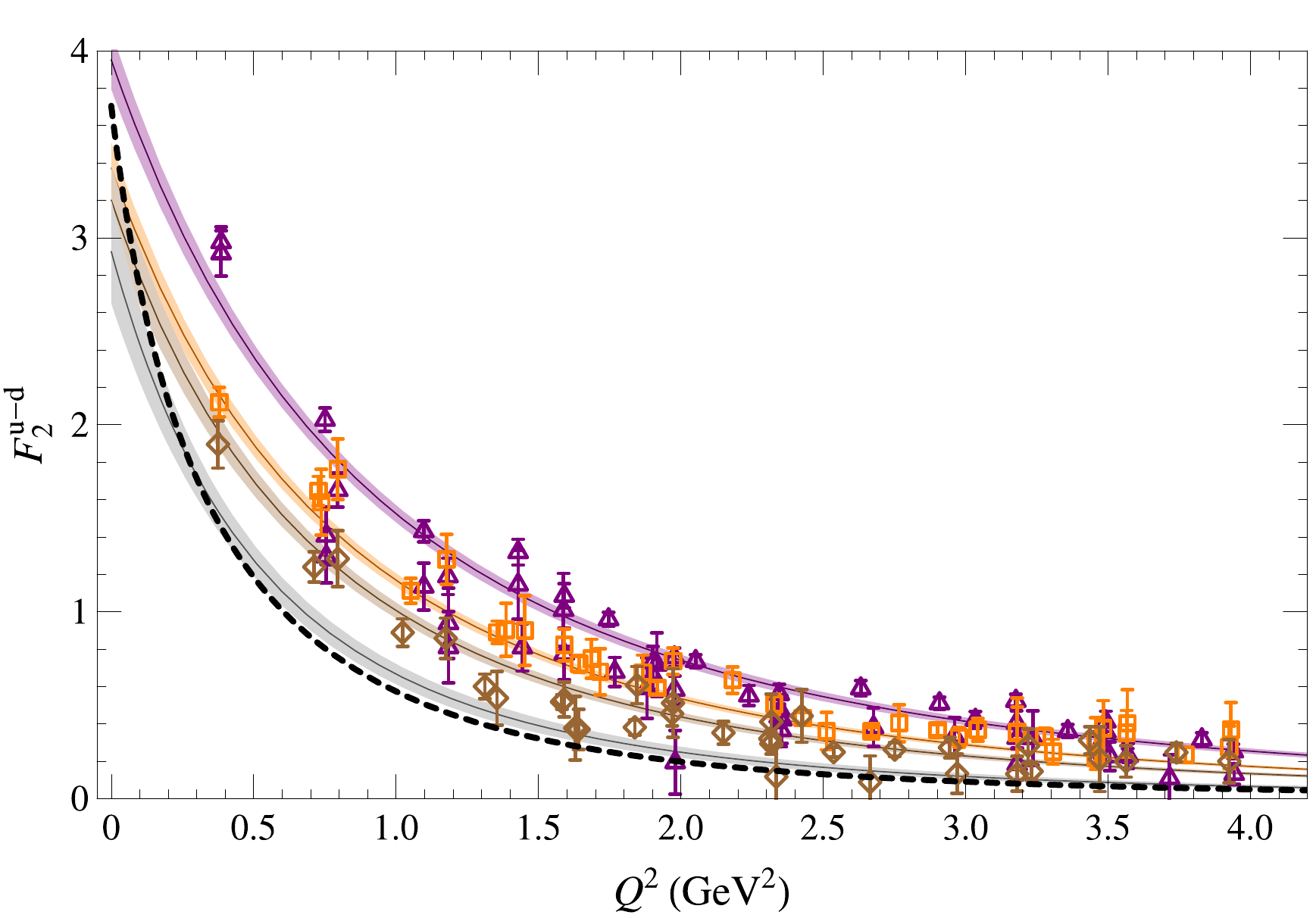}
\end{center}
\caption{\label{fig:F12}
Isovector Dirac (left) and Pauli (right) form-factor extrapolation using $N_f=2+1$ calculations in this work. The shaded bands show the simultaneous fit to the $Q^2$ dependence of each pion mass points (875, 580 and 450~MeV from top to bottom), where the color of the band matches the color of the points. The lowest gray band is the extrapolation to the physical pion mass.
}
\end{figure}

From the HSC ensembles~\cite{Edwards:2008ja,Lin:2008pr}, we use the $16^3 \times 128$ lattices with pion masses of 875, 580 and 450~MeV to calculate nucleon electromagnetic form factors. We extrapolate $F_{1,2}^{u,d}$ through a simple parametrization as used in interpolating experimental data:
\begin{eqnarray}
F_{1} =
\frac{a_0+\sum_{i=1}^{k-2}a_i\tau^i}{1+\sum_{i=1}^{k}b_i\tau^i} ;\;\;\
\left(\frac{F_{2}}{\kappa}\right) = \frac{1+\sum_{i=1}^{k-3}a_i\tau^i}{1+\sum_{i=1}^{k}b_i\tau^i}, \label{eq:F12-fitform}
\end{eqnarray}
where $\tau = Q^2/(4M^2)$. We simultaneously fit the $m_\pi$ and $Q^2$ (or $\tau$) dependence of the lattice data, expanding each $Q^2$-fit parameter in terms of the pion mass: $a_i=a_i^{(0)}+a_i^{(1)} m_\pi^2$. 
The right-hand side of Fig.~\ref{fig:F12} shows an example of the extrapolations on the isovector Dirac form factor in the dynamical ensembles. The $\chi^2/{\rm dof}$ for $F_{1,2}^v$ are 1.4 and 2.0, respectively. 
The lowest line/band represents the extrapolated form factors at the physical pion mass. 
We find good agreement between the Dirac form factor extrapolation at the physical pion mass and the experimental values; that is, the curvature (which is proportional to the product of squared nucleon mass and the mean-square charge radius) as a function of $Q^2$ fits nicely with interpolating forms derived from experimental data.
However, the Pauli form factor at physical pion mass suffers slightly larger $\chi^2/{\rm dof}$, mostly due to the failure to reproduce the (anomalous) magnetic moment, which would require smaller-momentum data (${}< 500$~MeV) that are not accessible with the current boundary conditions. At large momentum transfer, it becomes more consistent with the experimental form.

\begin{figure}
\begin{center}
\includegraphics[width=0.45\textwidth]{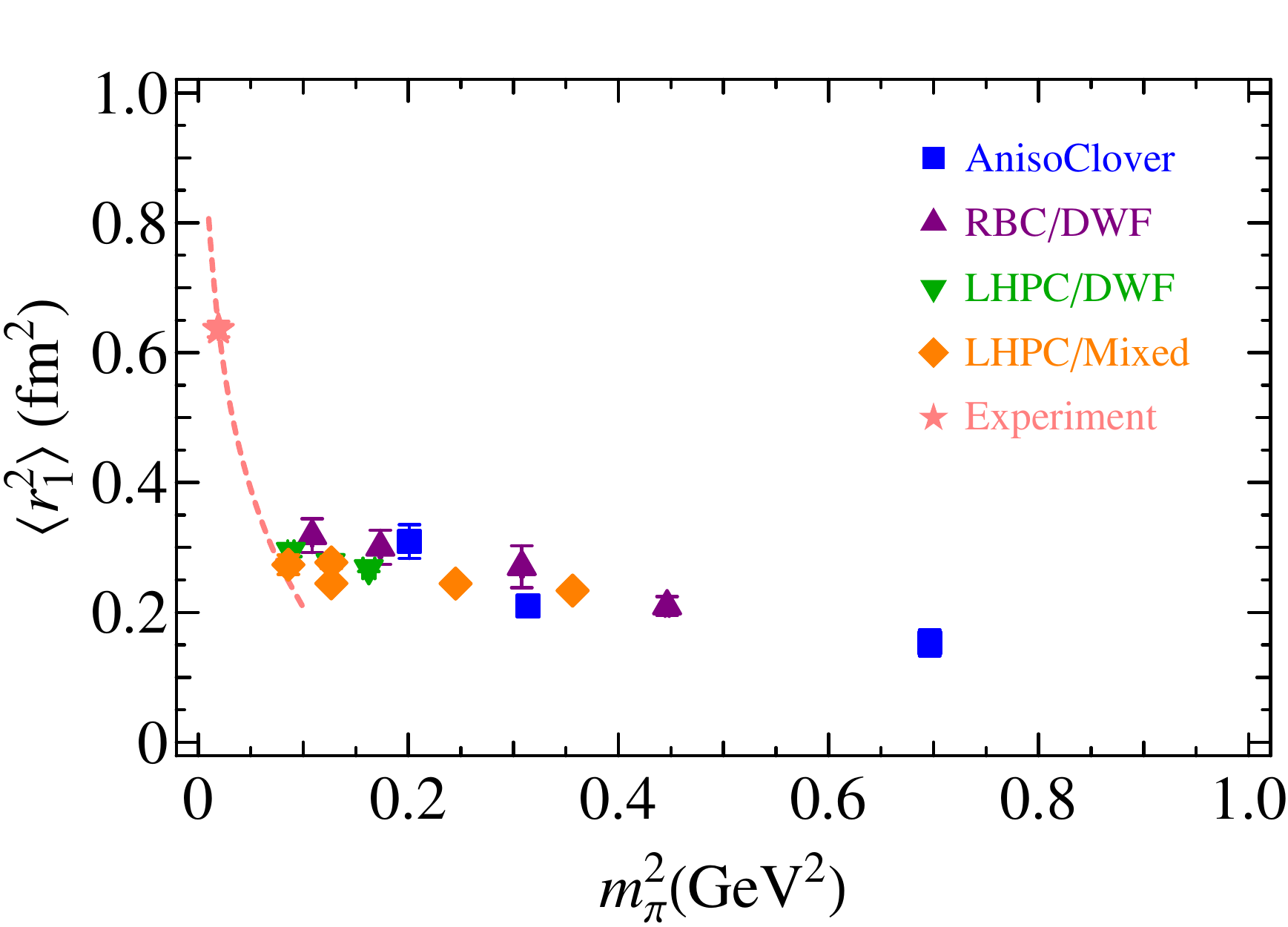}
\includegraphics[width=0.45\textwidth]{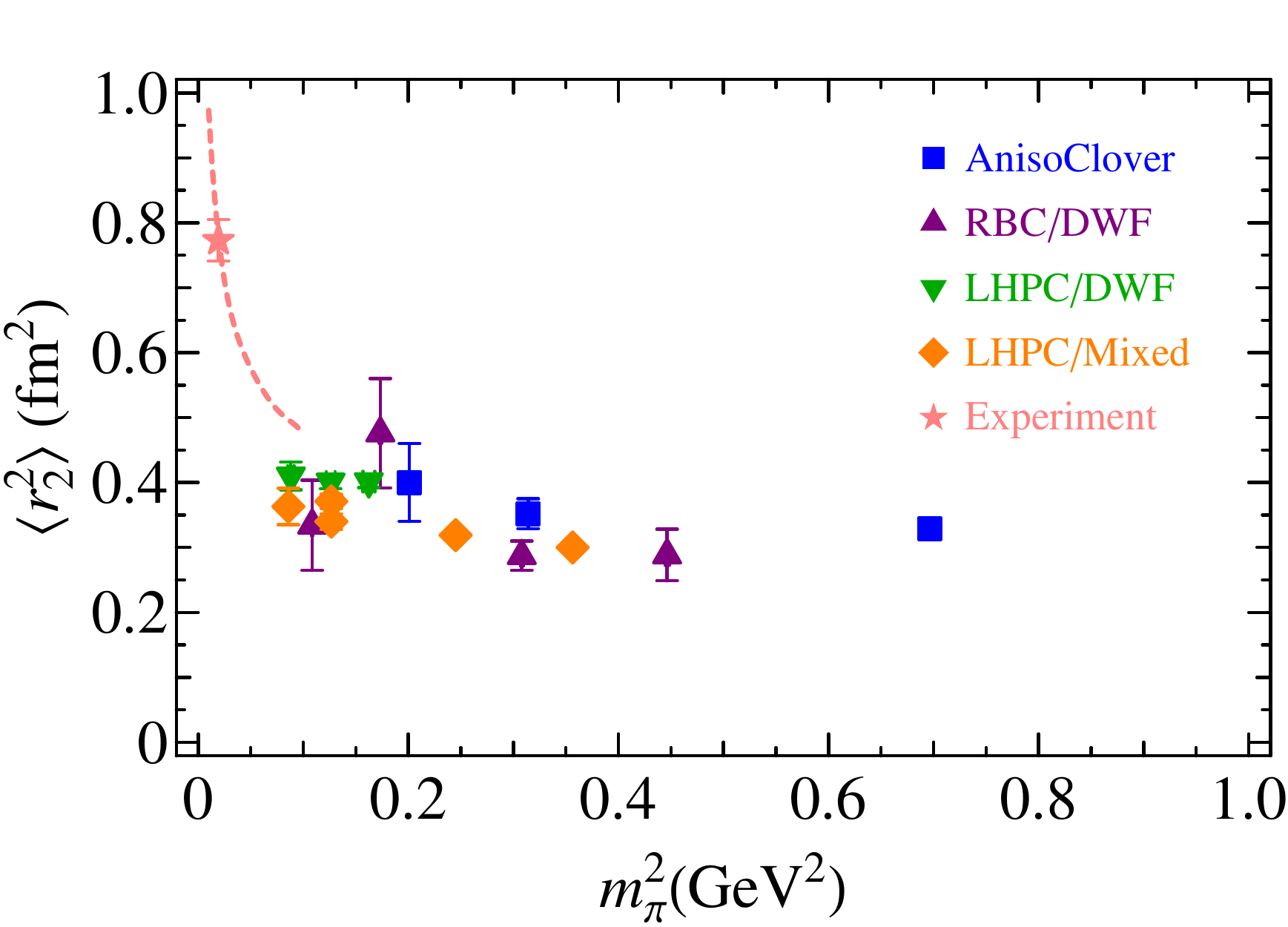}
\includegraphics[height=.21\textheight]{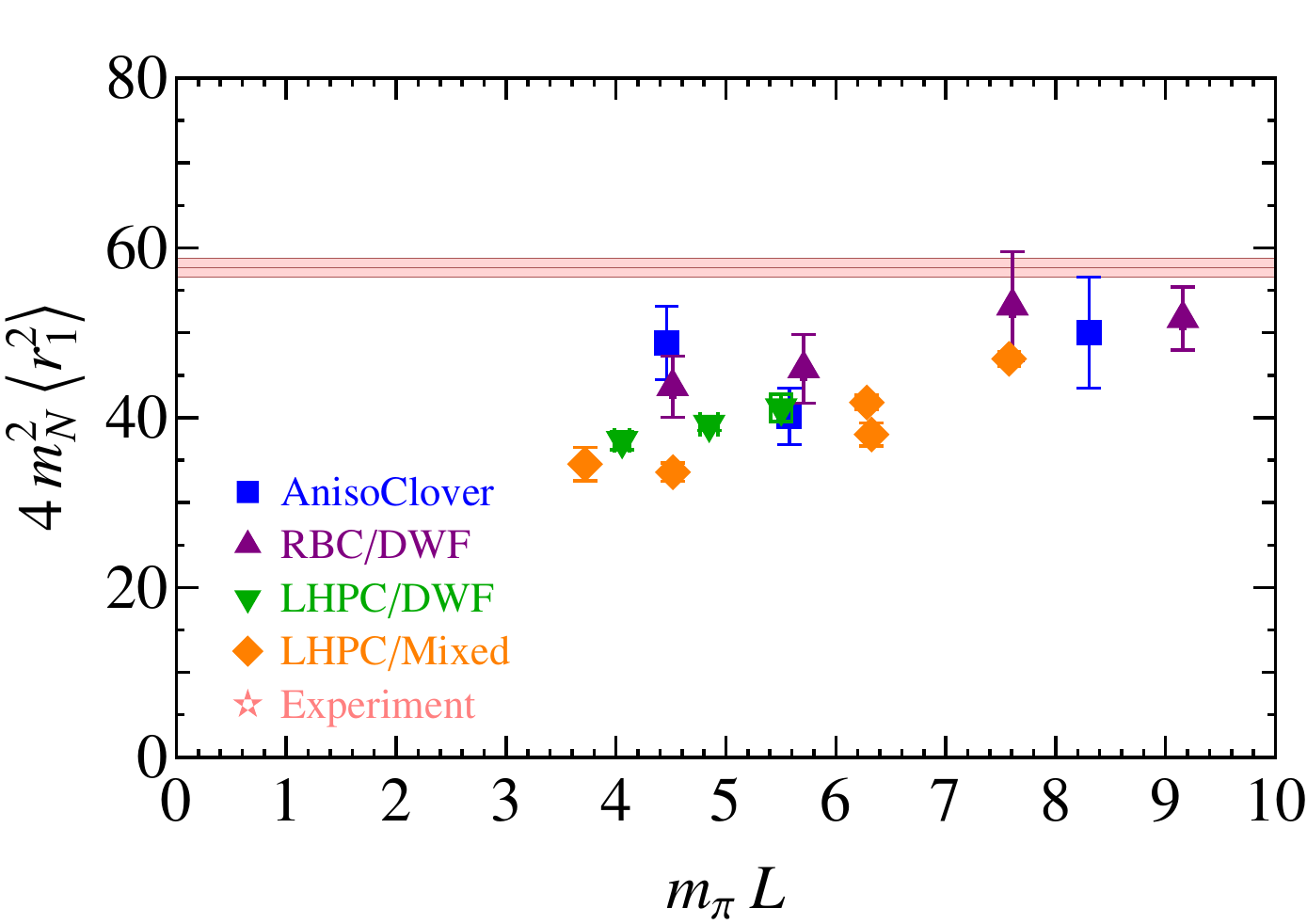}
\includegraphics[width=0.45\textwidth]{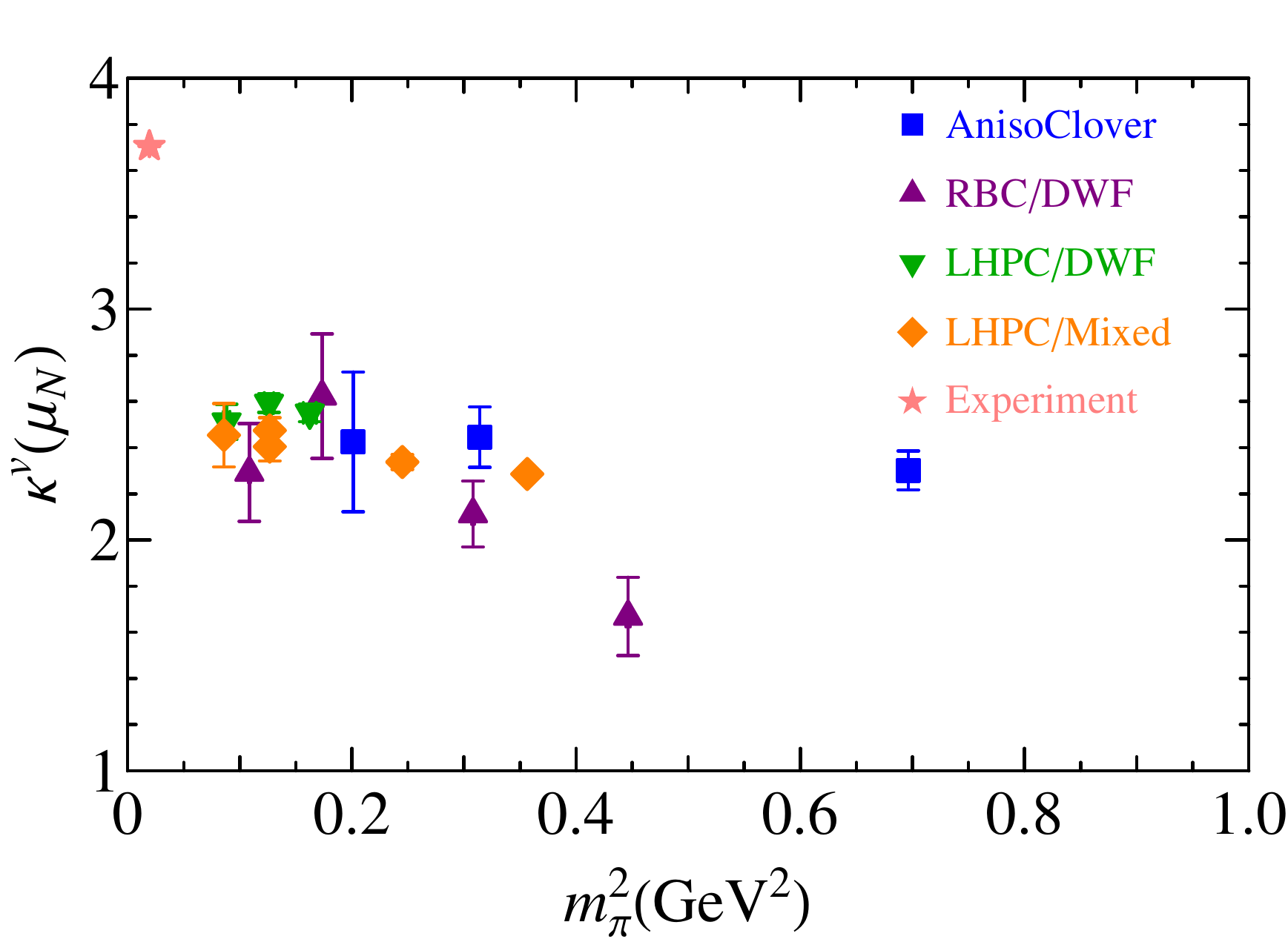}
\end{center}
\caption{Summary of the isovector Dirac (upper left) and Pauli (upper right) mean-squared radii from all currently existing $N_f=2+1$ nucleon electromagnetic form-factor calculations~\cite{:2010jn,Yamazaki:2009zq,Syritsyn:2009mx,Lin:2008mr,Lin:2010fv}. The dashed line indicates the leading-order HBXPT prediction.
(lower left) Summary of the dimensionless products of Dirac radii and nucleon mass from each $N_f=2+1$ lattice calculation, as a function of $m_\pi L$. The horizontal line indicates the experimental values.
(lower right) Summary of the normalized isovector anomalous magnetic moments from all currently existing $N_f=2+1$ nucleon electromagnetic form-factor calculations~\cite{:2010jn,Yamazaki:2009zq,Syritsyn:2009mx,Lin:2008mr,Lin:2010fv}, including this work.\label{fig:all21-rv2}}
\end{figure}

To compare with other existing form-factor calculations, we can examine quantities in the small-$Q^2$ region, such as the isovector Dirac and Pauli mean-squared radii, from the isovector electric form factors $F_{1,2}^v$ via
\begin{eqnarray}\label{eq:GEradii}
\langle r_{1,2}^2\rangle &=& (-6)\left.\frac{d}{dQ^2}\left(\frac{F_{1,2}^v(Q^2)}{F_{1,2}^v(0)}\right)\right|_{Q^2=0}.
\end{eqnarray}
Most groups have studied radii with the $Q^2$ dependence over ranges 0.5--2.0~GeV$^2$ and found the extracted radii to be independent (within the statistical error bars) of $Q^2$ choice~\cite{:2010jn,Syritsyn:2009mx,Lin:2008mr}. 
The Dirac and Pauli mean-squared radii from the dynamical ensembles are summarized in the upper row of Fig.~\ref{fig:all21-rv2} along with other $N_f=2+1$ lattice calculations and lowest-order heavy-baryon chiral perturbation theory (HBXPT) using experimental inputs~\cite{
Bernard:1998gv}. A dimensionless quantities, $4m_N^2\langle r_1^2\rangle$ versus $m_\pi L$ is also shown in Fig.~\ref{fig:all21-rv2}, in case the discrepancy is caused by setting the lattice spacing. Our results are nicely in agreement with isotropic $N_f=2+1$ calculations having various sea and valence fermion actions; this demonstrates the universality of the lattice-QCD calculations. However, both mean-squared radii are a factor of 2 smaller than the experimental values, and HBXPT suggests a rapid increase of the radii for pion masses between 250 and 140~MeV; near-future calculations will be able to check this out.
The (anomalous) magnetic moment, defined as $\kappa_N = F_2^N(Q^2=0)$, is also shown in the right-most figure in Fig.~\ref{fig:all21-rv2}.

Since the elastic form factors contain information about the spatial structure of the nucleon, we can convert our data into a description of its densities in the impact plane.
Having form factors at larger momentum transfer can greatly influence our understanding of the short-distance hadron structure.
Due to the relativistic effects of the transferred momentum on the wavefunction of the nucleon, we cannot use a simple three-dimensional Fourier transformation without some recourse to models. Instead, we use the model-independent formulation of Ref.~\cite{Miller:2007uy,Miller:2010nz} in terms of densities in a two-dimensional plane transverse to an infinite-momentum boost.
The transverse charge density ($\rho$) 
is defined as the Fourier transform of the form factor in such a plane: 
\begin{equation}
\rho(\mathbf{b}) \equiv \int\!\frac{d^2\mathbf{q}}{(2\pi)^2}F_1(\mathbf{q}^2)e^{i\mathbf{q}\cdot\mathbf{b}},
\end{equation}
where bold vectors $\mathbf{b}$ and $\mathbf{q}$ lie in the transverse plane. Equivalently,
\begin{equation}
\label{eq:rho}
\rho(b) = \int_0^\infty\!\frac{Q\,dQ}{2\pi}J_0(bQ)F_1(Q^2),
\end{equation}
for scalar $b$, where $J_0$ is a Bessel function. We can perform this integral numerically, using the $F_1(Q^2)$ obtained by extrapolating our fit form to the physical pion mass. (Similarly, one can examine the magnetic density ($\rho_M$) by integrating over $J_1$ Bessel function and $F_2$ form factor.)
In Ref.~\cite{Lin:2010fv}, we demonstrate the importance of the high-$Q^2$ form factors
by restricting the data set to the region $Q^2 < 2$~GeV$^2$ (which is the upper limit of transfer momentum for many lattice-QCD calculations) and comparing the result to the density obtained from using all the available $Q^2$. The impact is significant in the central core, as we anticipated; omitting information about large transfer momenta results in a deviation in the density around 25\%. (See Fig.~13 in Ref.~\cite{Lin:2010fv}.) Transverse charge density and magnetic density in infinite-momentum frame are shown in Figs.~14--16 in Ref.~\cite{Lin:2010fv}.

\begin{figure}
\begin{center}
\includegraphics[width=0.48\columnwidth]{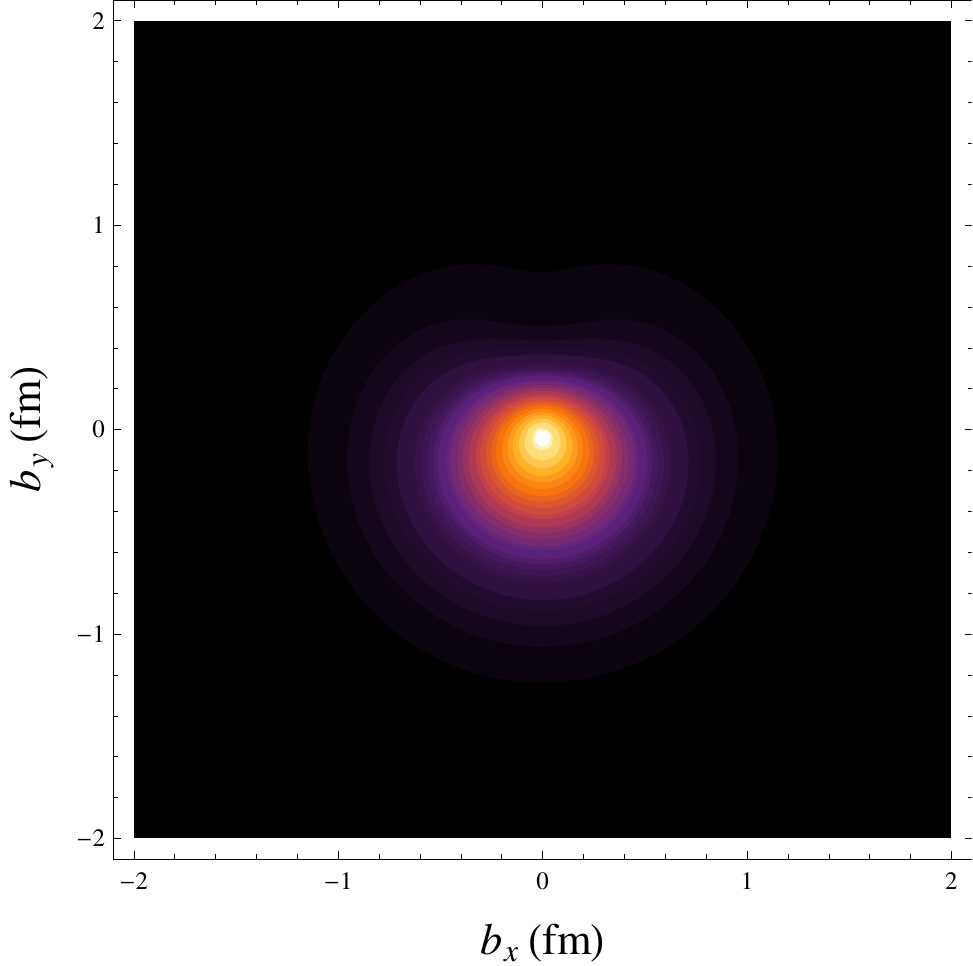}
\includegraphics[width=0.48\columnwidth]{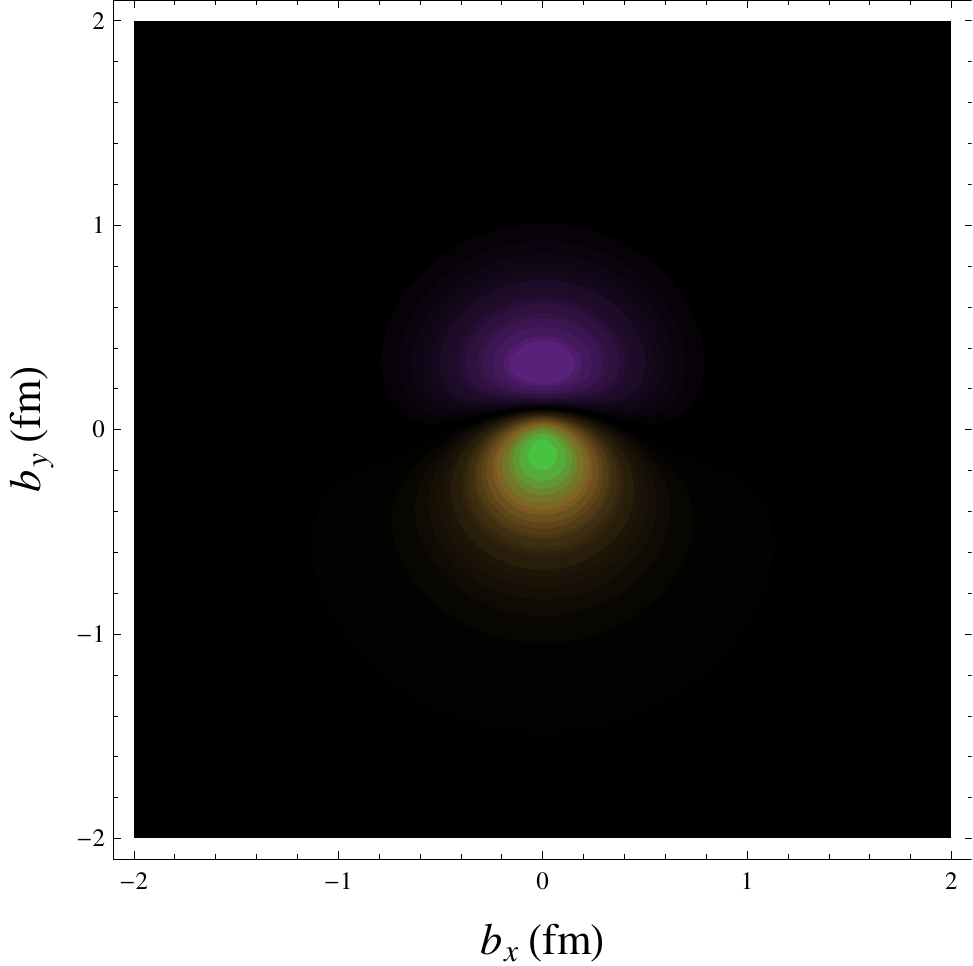}
\includegraphics[width=0.48\columnwidth]{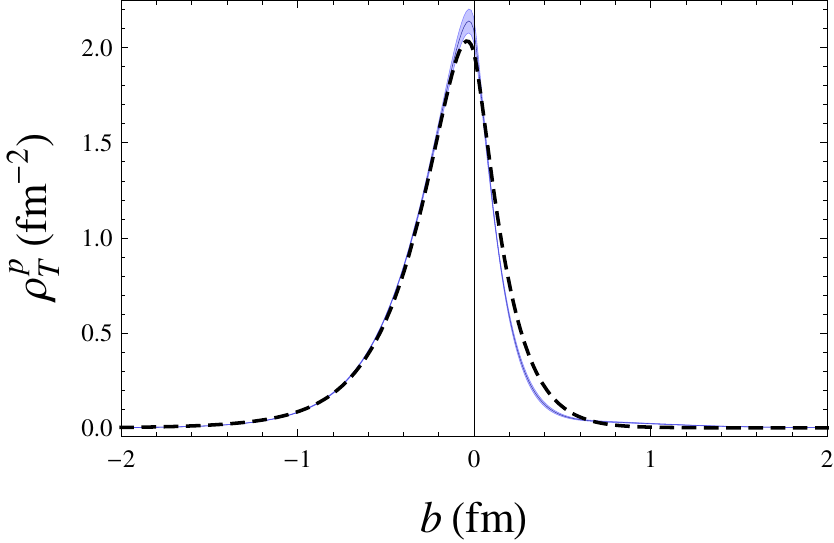}
\includegraphics[width=0.48\columnwidth]{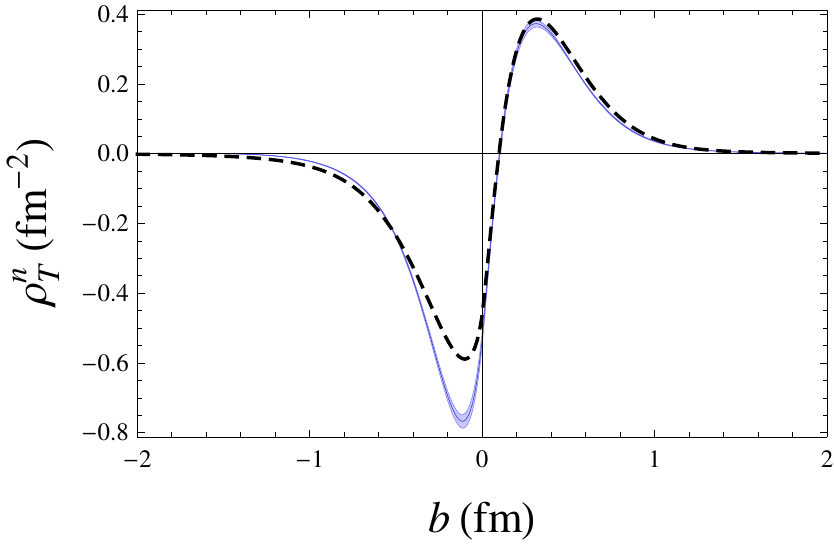}
\end{center}
\caption{The transverse charge densities in a polarized nucleon. The left- (right-) column figures present the results of proton (neutron) with $b$ (lower) and $b_{x,y}$ (upper) in fm. 
In the upper-row figures, black indicates near-zero values, and purple, orange and white are increasingly positive. For the right figure, the same color-scale applies, slightly negative values are gold and the most negative are green.
In the lower-row figures, the blue band indicates the densities from our lattice 2+1-flavor calculation, while the dashed line is a parametrization interpolating and extrapolating from available experimental data~\cite{Arrington:2007ux,Kelly:2004hm}. (Note that the neutron experimental data is extrapolated to larger $Q^2$ region). 
}
\label{fig:F-density}
\end{figure}

We can further look at the transverse charge densities in a polarized nucleon~\cite{Carlson:2007xd} via
\begin{equation}
\label{eq:rho_T}
\rho_T(b) = \rho + sin (\phi) \int_0^\infty\!\frac{Q^2\,dQ}{2\pi M_N}J_1(bQ)F_2(Q^2),
\end{equation}
where $\rho(b)$ is obtained from Eq.~\ref{eq:rho}. Results for the deuteron~\cite{Carlson:2008zc} and nucleon-Roper transition~\cite{Tiator:2008kd} have also been studied with experimental data.
Figure~\ref{fig:F-density} shows the results for the proton (left column) and neutron (right column) in one-dimension (lower row) and the two-dimensional impact plane (upper row) using our lattice inputs. 
Our lattice proton density is similar to the experimental one, while the neutron has more significant deviations. This is likely due to the lack of disconnected diagrams in the form factors, which are closer to the magnitude of a typical neutron form factor but smaller than the proton one. Also experimentally, the neutron electromagnetic form factors $G_E^n$ only known to 1.5~GeV$^2$; 
thus a majority of the form-factor inputs are based on extrapolation to larger $Q^2$ region. Note that the asymmetry in the distribution for a polarized nucleon is due to the relativistic effect of boosting the magnetic moment of the baryon. This induces an electric dipole moment that shifts the charge distribution.

\subsection{Nucleon Axial Form Factors}
\begin{figure}
\begin{center}
\includegraphics[width=0.48\columnwidth]{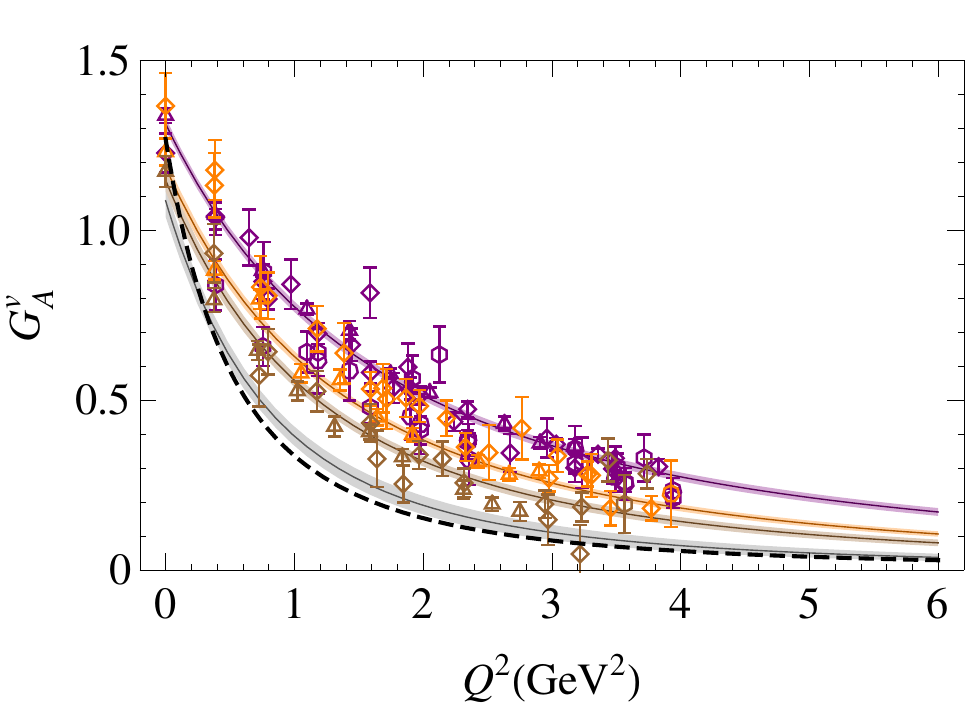}
\includegraphics[width=0.48\columnwidth]{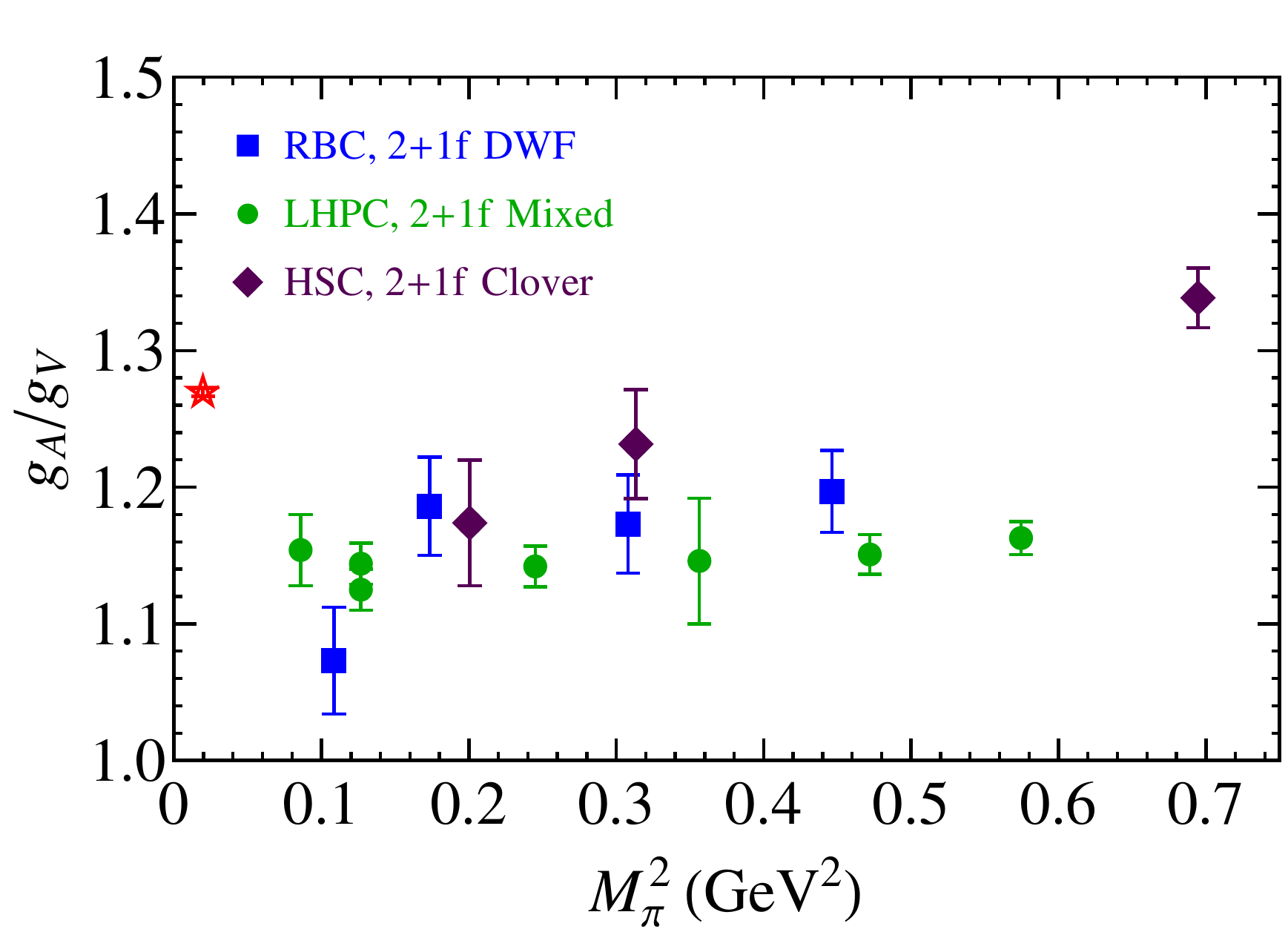}  
\includegraphics[width=0.48\columnwidth]{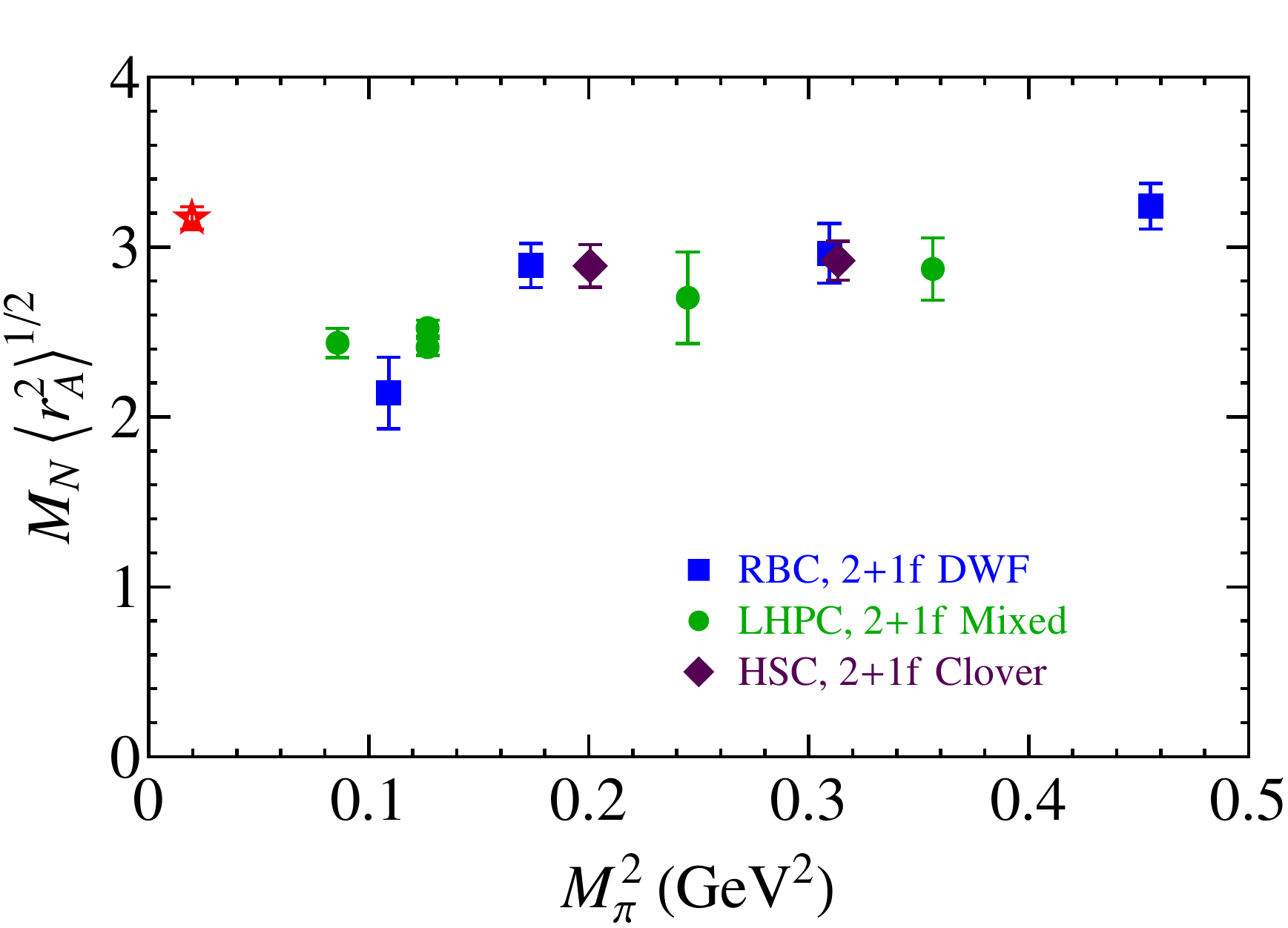}
\includegraphics[width=0.48\columnwidth]{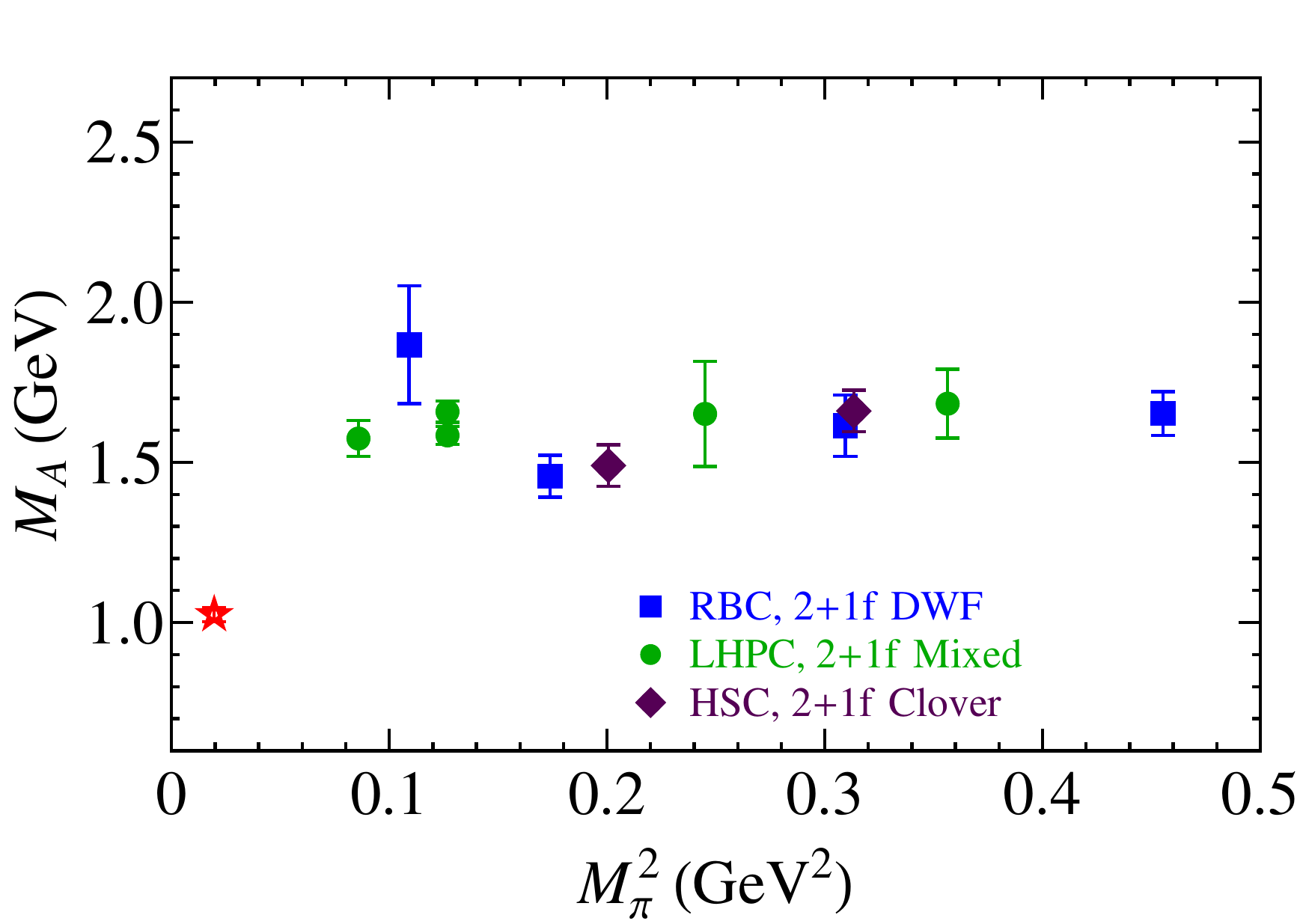}  
\end{center}
\caption{\label{fig:GA}
(upper left) Nucleon isovector axial form factors using 3 pion masses at 875, 580 and 450~MeV. The lowest gray band is the extrapolation to the physical pion mass. The dashed line is a dipole form using $M_A=1.03(2)$~GeV extending beyond the $Q^2$ region of the available experimental data. 
$g_A$ (upper right) and $M_N \sqrt{r_A^2}$ (lower left) obtained from the same ensembles, and comparisons with previous $N_f=2+1$ results. 
(lower right) Polarized distribution in a longitudinally polarized neutron.
}
\end{figure}

A similar approach can be applied to nucleon axial form factors, where experiments mostly derive results from neutron beta decay or pion form factors where various theoretical models predict a wide range of possibilities for the large-$Q^2$ region. 
The upper-left graph of Fig.~\ref{fig:GA} shows the preliminary results for the isovector axial form factors using ensembles with 3 different pion masses. The data are simultaneously extrapolated in pion mass and $Q^2$ (as were the EM form factors). The lowest gray band is the result at physical pion mass, and the dashed line is the dipole form with best fit to the experimental data ($M_A = 1.026(21)$~GeV~\cite{Bernard:2001rs}).\footnote{The $M_A$ value used here is obtained from a weighted average of $M_A$ from (quasi)elastic neutrino and antineutrino scattering experiments only. Ref.~\cite{Bernard:2001rs} also analyzed weighted values from charged pion electroproduction experiments and obtained $M_A = 1.069(16)$~GeV. Recent MiniBooNE analysis suggests the $M_A$ could get as high as 1.3~GeV, while other recent experiments using neutrino scattering off various targets give a range 1.14--1.26~GeV. Here we took an earlier $M_A$ from Ref.~\cite{Bernard:2001rs} as a guideline to compare with our lattice results while the experimental determination of $M_A$ remains unclear.} 
The nucleon axial coupling constants are obtained taking $G_A(Q^2=0)$ (purple diamonds) which is consistent with other $N_f=2+1$ results (RBC/UKQCD~\cite{Yamazaki:2008py} and LHPC~\cite{Bratt:2010jn}, statistical errorbar only) as functions of $m_\pi^2$, as shown in the upper-right graph of Fig.~\ref{fig:GA}.
The $\langle r_A^2\rangle$ is obtained from the curvature of the axial form factor, similar to the Dirac and Pauli formulation in Eq.~\ref{eq:GEradii}. A dimensionless combination with the lattice nucleon mass is plotted as a function of $m_\pi^2$ in the lower-left graphic of Fig.~\ref{fig:GA}; again, we see consistency with other $N_f=2+1$ results. In the large-$m_\pi$ region, we find the lattice products to be consistent with the experimental ones, while at lower $m_\pi$ the deviation becomes more significant. However, these points have $m_\pi L < 5$; it may be the finite-size effect dominates in this quantity. Further studies are needed resolve this issue. Finally, the nucleon axial mass $M_A$ from our calculation and other $N_f=2+1$ ones are displayed in the lower-right of Fig.~\ref{fig:GA}.

\begin{figure}
\begin{center}
\includegraphics[width=0.48\columnwidth]{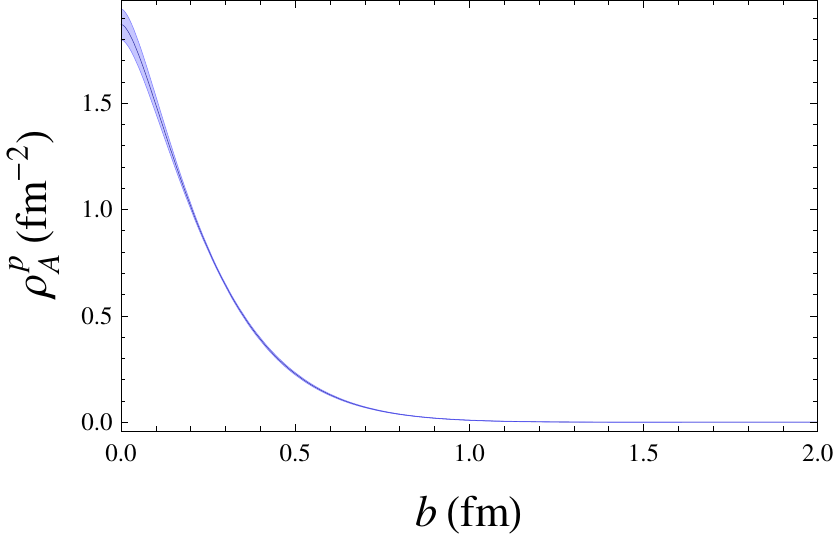}  
\includegraphics[width=0.48\columnwidth]{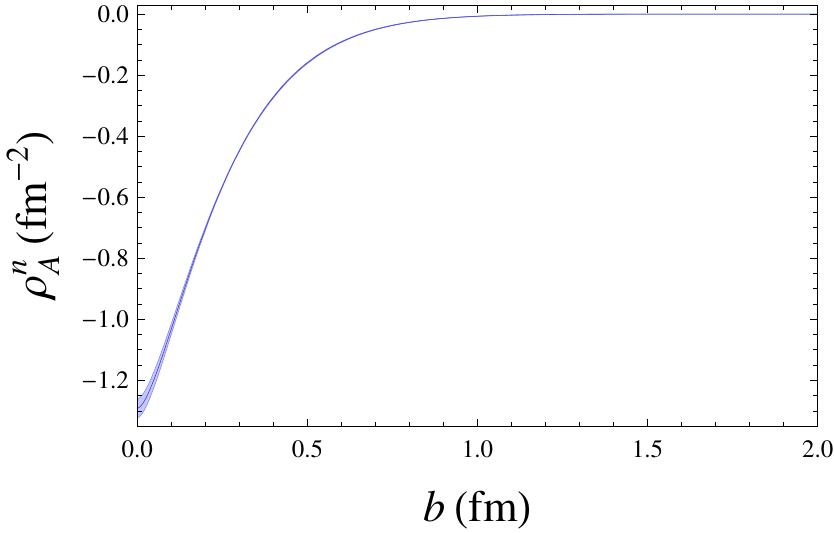}\\
\includegraphics[width=0.48\columnwidth]{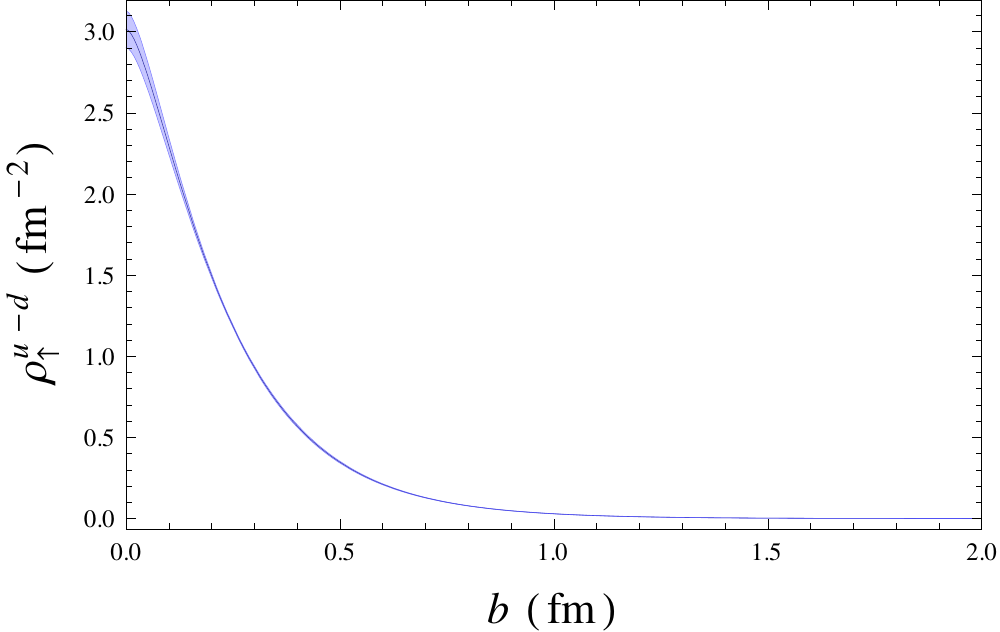}
\includegraphics[width=0.48\columnwidth]{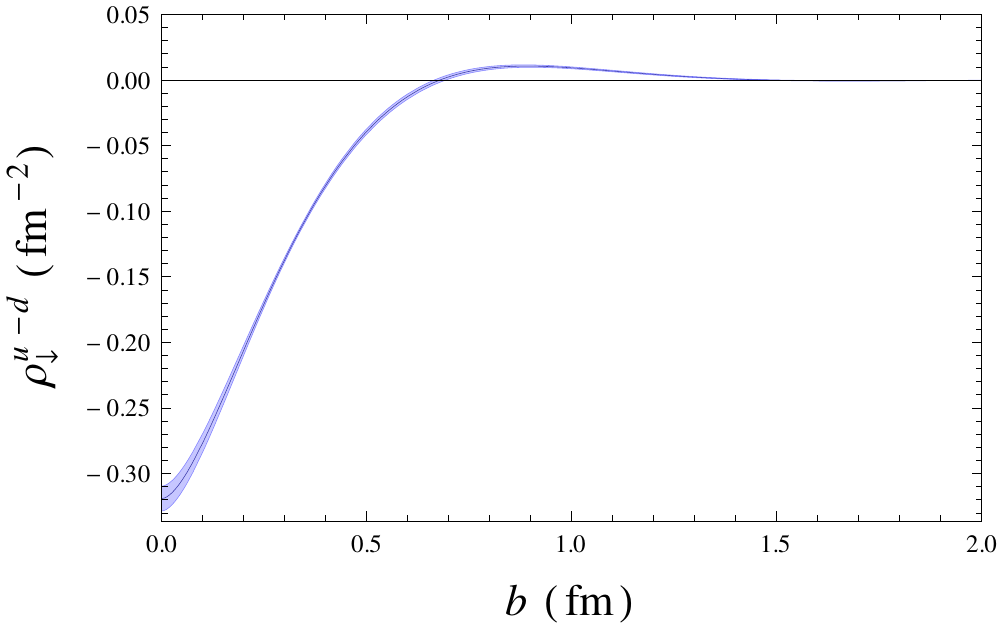}
\end{center}
\caption{\label{fig:helicity}
(upper row) Polarized distribution in a longitudinally polarized proton (left) and neutron (right).
(lower row) Isovector helicity distribution of nucleon.
}
\end{figure}
Unlike the nucleon electromagnetic (Dirac and Pauli) form factors, little is known about the nucleon axial form factors at large $Q^2$ from experiments. With the large-$Q^2$ nucleon axial form factors, we can study the polarized distribution in a longitudinally polarized proton (upper left in Fig.~\ref{fig:helicity}) and neutron (upper right in Fig.~\ref{fig:helicity}) for the first time via a similar three-dimensional Fourier transformation as Eq.~\ref{eq:rho}.
The magnitudes of the proton and neutron densities monotonically decrease away from the center of the impact plane. Unlike the transverse charge density in the unpolarized nucleon, the neutron density does not change sign.
Combining with the Dirac and Pauli form factors, we can look at the helicity distribution of nucleon in the impact plane~\cite{Burkardt:2002hr}; the isovector nucleon results are shown in the lower row of Fig.~\ref{fig:helicity}.

\subsection{Pion Form Factors}

\begin{figure}
\begin{center}
\includegraphics[width=0.28\textheight,angle=90]{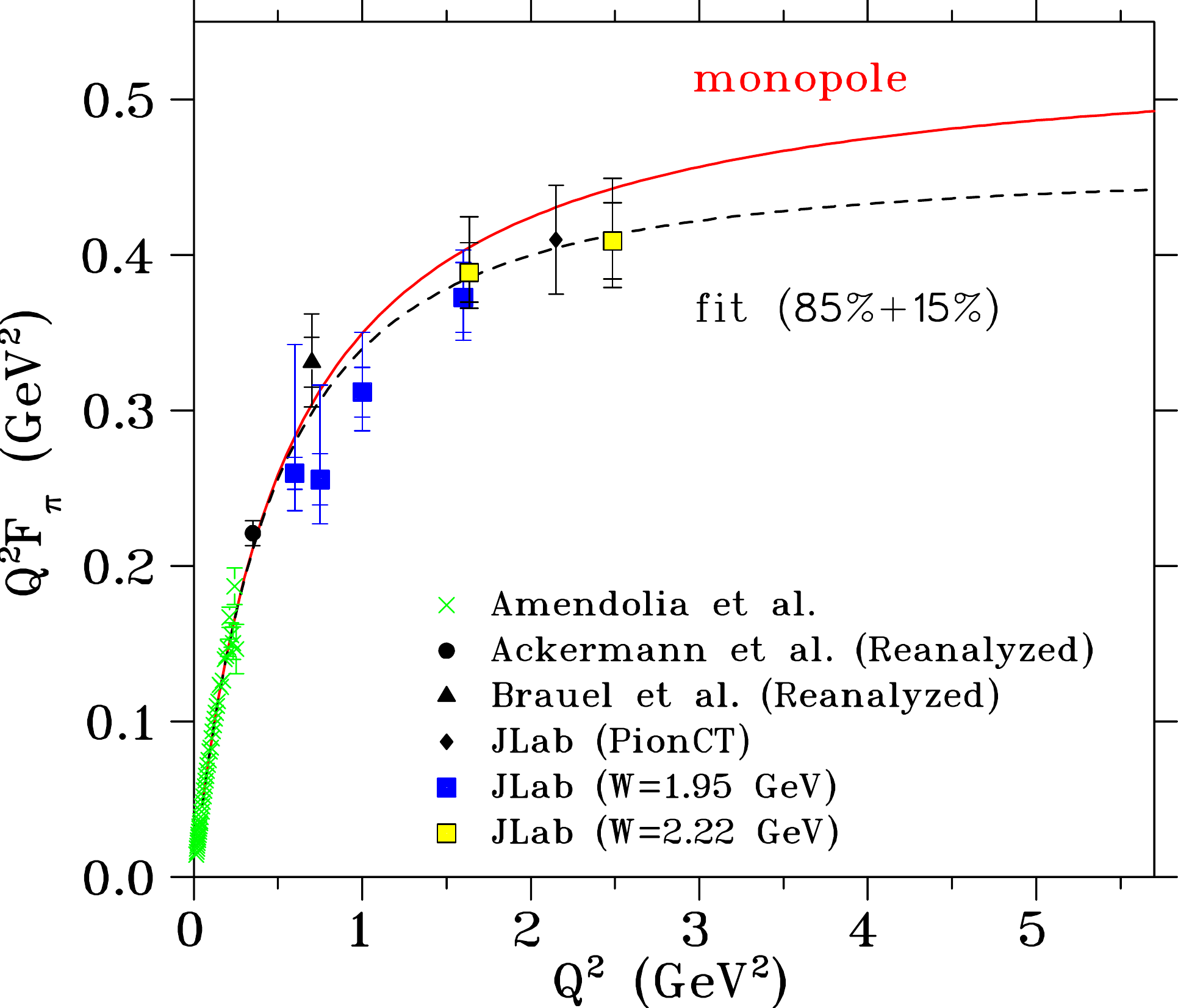}
\includegraphics[width=0.28\textheight,angle=90]{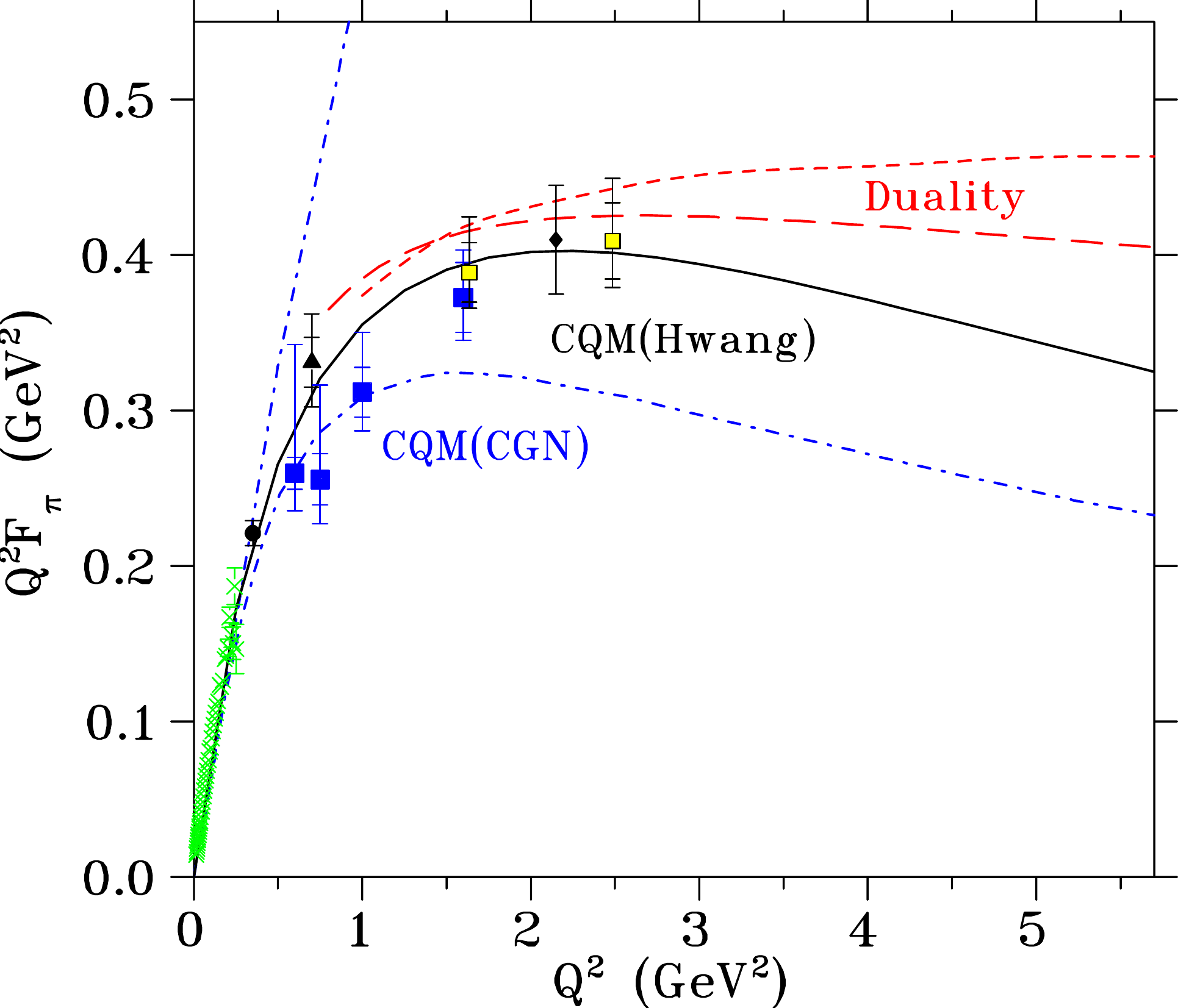}
\end{center}
\caption{\label{fig:exp-pion}
(left) Experimental data points (seen the legend in the figure) and the monopole (red) and a mixed monopole-dipole (dashed) fit to the data. 
(Right) An example form-factor calculation from the constituent-quark model (CQM); different lines (theoretical projections) indicate different quark wavefunctions within the same model. Figures taken from Ref.~\cite{Huber:2008id}.
}
\end{figure}

Recent high-precision pion electroproduction data from Jefferson Lab using L/T-separated cross sections provide precise pion form-factor measurements up to 2.45~GeV$^2$~\cite{Huber:2008id}. A summary of all the experimental data is shown in the left-hand side of Fig.~\ref{fig:exp-pion}; different-color points indicate data from different experiments. See Ref.~\cite{Huber:2008id} for more details. The solid red line is the fit using the monopole form, and the dashed line is the fit from an 85\% monopole + 15\% dipole form to the experimental data. 
Many theoretical models ranging from various quark models to holographic QCD provide diverse and disagreeing predictions for the intermediate and high-$Q^2$ dependence of the pion form factors. Even within the framework of constituent-quark model, differences in the treatment of the quark wavefunctions result in very different pion form factors; see the right-hand side of Fig.~\ref{fig:exp-pion} (taken from Ref.~\cite{Huber:2008id}). Diverse behaviors are obtained from various theoretical calculations. Future precision experiments will challenge these models even more, aiming at $Q^2$ through 6~GeV$^2$ and higher.

\begin{figure}
\begin{center}
\includegraphics[width=0.48\columnwidth]{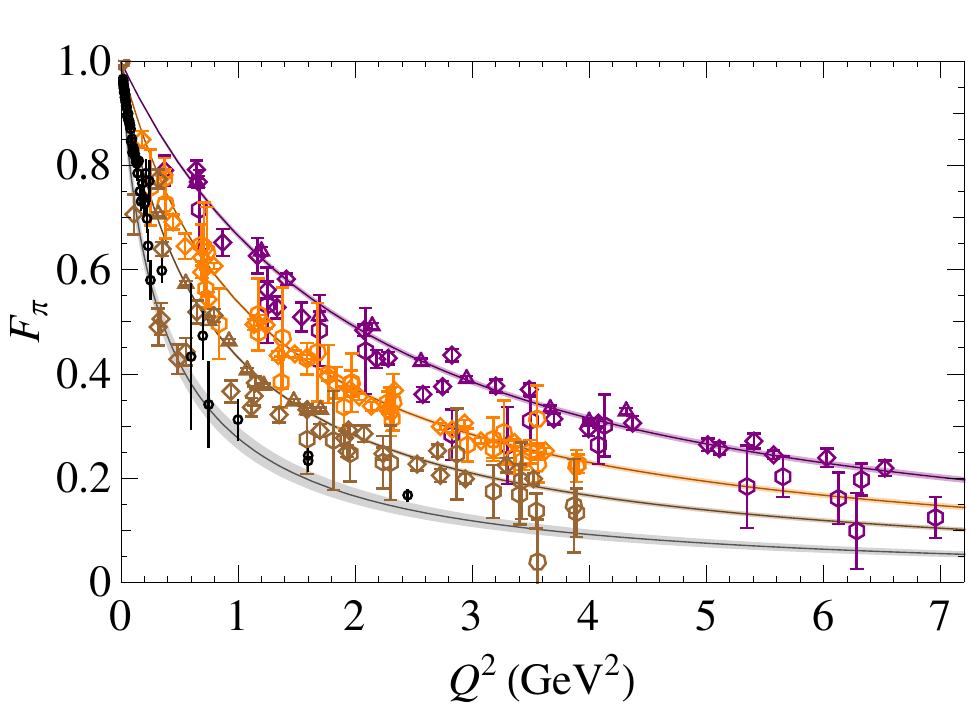}
\includegraphics[width=0.48\columnwidth]{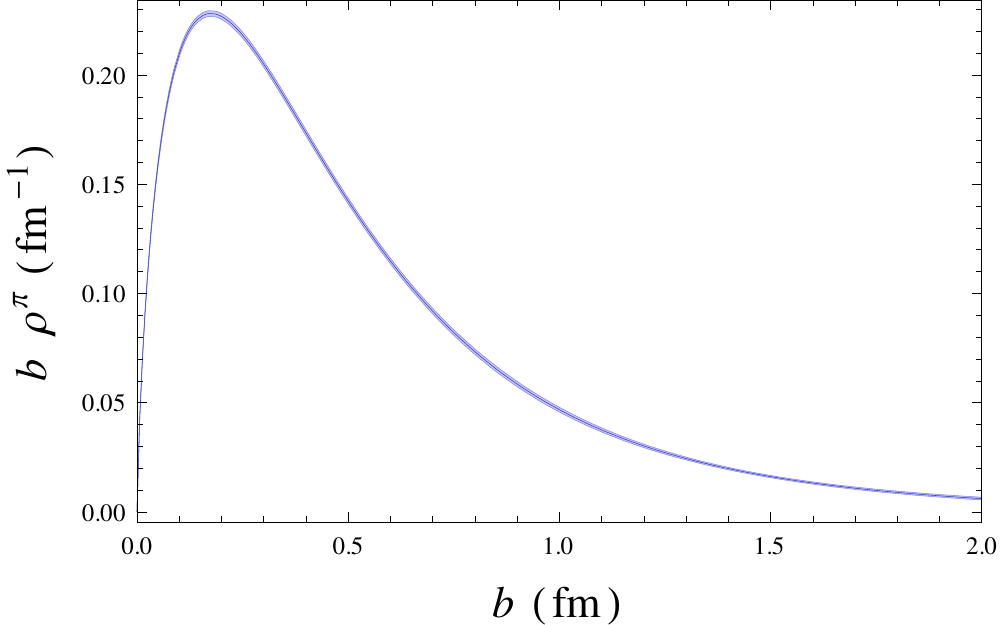}
\end{center}
\caption{\label{fig:pion}
(left) Pion form factor with multiple pion masses at 580, 875, 1350~MeV. The experimental points are shown as black circles while the lowest gray band is the extrapolation to the physical pion mass using our lattice points.
(right) $b\rho_\pi$ from our lattice pion form-factor data.
}
\end{figure}

We apply the same methodology as the nucleon large-$Q^2$ form-factor calculation to the pion form factor to understand and make predictions for upcoming experiments.
The right-hand side of Fig.~\ref{fig:pion} shows the preliminary results for our dynamical ($N_f=2+1$) pion form factors with 580, 875, 1350~MeV pion masses with $Q^2$ reaching nearly 7~GeV$^2$ for the highest-mass ensemble. The extrapolated form factor at the physical pion mass shows reasonable agreement with JLab precision measurements. Future attempts will focus on decreasing the pion masses and exploring $Q^2$-dependence of pion form factors for yet higher $Q^2$.

We can further use our pion form factor data to study the three-dimensional infinite-momentum--frame spatial charge density of the pion~\cite{Miller:2009qu}:
\begin{equation}
\label{eq:rho}
\rho_\pi(b) = \int_0^\infty\!\frac{Q\,dQ}{2\pi}J_0(bQ)F_\pi(Q^2).
\end{equation}
$\rho_\pi$ and $b\rho_\pi$ were first studied using experimental data up to 2.45~GeV$^2$ in Ref.~\cite{Miller:2009qu}, along with a few perspectives from perturbative QCD, QCD sum rules, holographic QCD, and the Nambu--Jona-Lasinio model.
Perturbative QCD predicts the asymptotic behavior for the pion form factor at large-$Q^2$ region, \begin{equation}
\label{eq:FpiPQCD}
\lim_{Q^2\rightarrow \inf} F_\pi(Q^2) =16 \pi \alpha_s (Q^2)f_\pi^2/Q^2 .
\end{equation}
This leads to a singular central charge density for the pion, which is unusual in nature.

The right-hand side of Fig.~\ref{fig:pion} shows $b\rho_\pi$ using our calculation of lattice pion form factors. Here we use a two-pole form to fit our lattice data. The singularity at $b=0$ is highly dependent on the $Q^2$ behavior. However, it is not clear from the current precision which $Q^2$ form is better suited to describe our data. Further studies are needed to answer this question.

\begin{figure}
\begin{center}
\includegraphics[width=0.48\columnwidth]{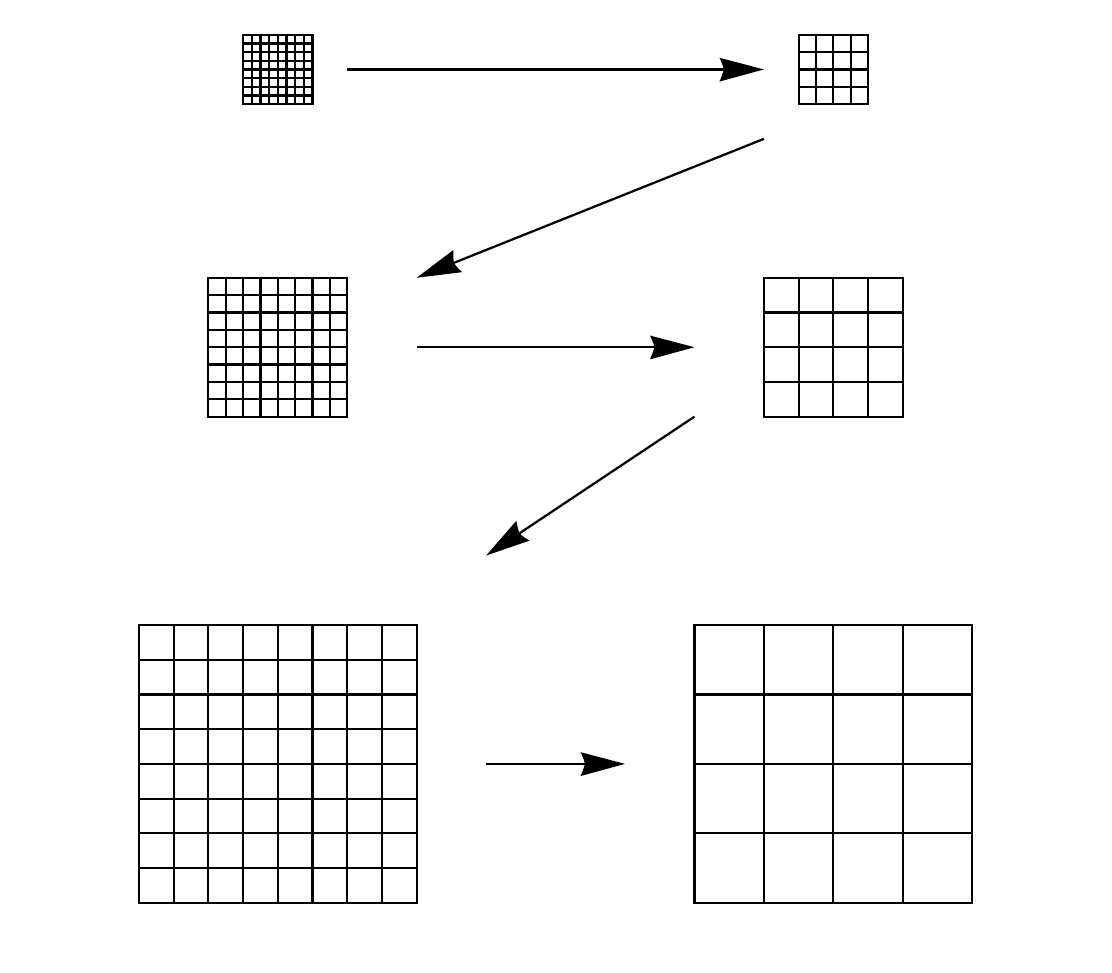}
\end{center}
\caption{Proposed idea for extending the current $Q^2$ regime to higher ones}
\label{fig:stepscale}
\end{figure}

\section{Conclusion and Outlook}

Starting from the earlier exploratory study of large momentum transfer on the lattice for nucleon Dirac and Pauli form factors, we employed a strategy to overcome noise in the conventional method at large $Q^2$. The idea is to include operators that couple to high-momentum and excited states and explicitly analyze excited states to get better ground-state signal.
We demonstrate that one can extend form-factor calculations to higher momentum transfer with reasonable statistical error at a fixed lattice spacing. However, systematic error could become significant when one reaches a large value of $\left|n\right|$.
What we shown for heavier pions on anisotropic lattices can be easily extended to isotropic lattice or lighter-pion calculations.

We demonstrate that our analysis shows consistent (within systematics) results for low-$Q^2$ physical quantities, such as $\langle r_{1,2}^2 \rangle$, $\kappa$, $g_A$, $M_A$, comparing with RBC/UKQCD and LHPC's published $N_f=2+1$ results. We calculate larger-$Q^2$ form factors that these calculations lack.
With large $Q^2$, we can study the various densities of hadrons that depend on a wide range of momentum in form factors. In this proceeding, we have shown the results for transverse charge densities in unpolarized and polarized nucleons,
polarized distributions in longitudinally polarized nucleons, and isovector helicity distributions of nucleons. The transverse pion density is also calculated, but we need more precision to explore whether there is a singularity at the center of the pion.

Future work will include smaller source-sink separation to improve the signal; since excited-state analysis is included in our methodology, this is not an issue for our ground-state nucleon or pion form factors. It would be interesting to study the same form factor using multiple lattice spacings to study and reduce systematic errors at large-$Q^2$ due to lattice discretization error.

To reach even higher $Q^2$, we propose a numerical step-scaling calculation, as shown in Fig.~\ref{fig:stepscale}), On a small volume with very fine lattice spacing, we can easily reach high momentum.
By calculating the step-scaling function at overlapping momentum points (or interpolating) we can reduce the systematic error due to finite-volume or lattice-discretization artifacts. However, generating several volumes and tuning for multiple lattice spacing of dynamical $N_f=2+1$ lattices requires a large amount of computational resources; we hope such a proposal will become feasible as petascale computing facilities become available in the near future.

\section*{Acknowledgments}
The form factor data is taken using USQCD resources and UW Hyak cluster, NSF MRI PHY-0922770.
HWL is supported by the U.S. Dept. of Energy under grant No. DE-FG03-97ER4014. SDC is supported by DOE grant Nos. DE-FG02-91ER40676 and DE-FC02-06ER41440 and NSF grant No. 0749300.


\end{document}